\documentstyle[preprint,prb,aps]{revtex}
\begin{document}
\draft
\title{The quantum Hall effect and inter-edge state\\
tunneling within a barrier}
\author{B.L. Johnson,$^1$ A.S. Sachrajda,$^2$ G. Kirczenow,$^1$
Y. Feng,$^2$ R.P. Taylor,$^{2,3}$ L. Henning,$^2$ J. Wang,$^2$
P. Zawadzki,$^2$ and P.T. Coleridge$^2$} 

\address{$^1$Department of Physics, Simon Fraser University,
Burnaby, British Columbia, Canada V5A 1S6}

\address{$^2$Institute for Microstructural Sciences, 
National Research Council,
Ottawa, Ontario, Canada K1A 0R6}

\address{$^3$University of New South Wales, Kensington, NSW
2033, Australia}

\maketitle
\begin{abstract}
We have introduced a controllable 
nano-scale incursion into a 
potential barrier imposed across a two-dimensional 
electron gas, and report on the phenomena that we 
observe as the incursion develops. In the quantum Hall 
regime, the conductance of this system displays 
quantized plateaus, broad minima and oscillations. We 
explain these features and their evolution with electrostatic
potential geometry and magnetic field as a progression of 
current patterns formed by tunneling between edge 
and localized states within the barrier.
\end{abstract}

\narrowtext

Electronic transport in semiconductor systems of
reduced dimensionality has been a field of great
importance since the discovery of the quantum Hall
effect by von Klitzing, Dorda, and Pepper.\cite{vonK} They found
that the Hall conductance of a two dimensional electron
system at low temperatures is quantized, to
remarkable precision, in integer multiples of $e^2/h$.
Subsequently, Halperin,\cite{halp} Streda, et. al,\cite{streda}
Jain et. al.,\cite{jain} and
B\"uttiker\cite{butt1} proposed a theoretical picture in which the
quantum Hall effect is seen as a manifestation of
electrical conduction by magnetic edge states. These
states appear at the boundaries of a macroscopic Hall
bar in a strong magnetic field. Electrons that flow
through them are immune to backscattering\cite{butt1} in the
quantum Hall regime, which results in the quantization
of the Hall conductance and dissipationless transport.
However, the edge states can be selectively
backscattered if a potential barrier traverses the
entire sample, as was demonstrated by Haug et al,
\cite{haug,haug2} who
used this principle to elucidate the relationship
between the quantum Hall effect and its breakdown,
and the Landauer\cite{land} theory of one-dimensional
conduction. Experiments performed by Washburn et al.,\cite{wash} 
focused on the four-terminal conductance of a much smaller 
structure, a pair of narrow channels coupled by a barrier. 
These authors were also able to exploit backscattering 
from the barrier to study different transport
quantization regimes, but they found that oscillatory 
conductance fluctuations, apparently arising from
inhomogeneities in the electrostatic barrier potential, 
characterize the breakdown of quantization when it occurs.

The purpose of this paper is to report on an
experimental study of the role of such a potential inhomogeneity,
a depression introduced
intentionally and in a controlled way in a potential
barrier. We have been able to isolate the role of the 
internal structure of the depression in the transport 
problem experimentally, and propose a model that 
explains all of the phenomena that are observed 
as the potential depression gradually develops. We note that
although the goal of this work is different, the structure 
that we have constructed represents a different method
of measuring the transport properties of quantum dots, which
have been discussed elsewhere.\cite{vanwees}

One of the most common techniques for fabricating low-dimensional
devices is based on the surface split gate technique
developed by Thornton et al.\cite{thort} However, the geometries
used have been constrained by the conventional methods used to
contact surface gates. Recently, new techniques have been 
developed which for the first time enable isolated
submicron gates to be contacted \cite{camb,andy},
making the creation of a nanoscale depression
possible. A schematic of our experimental
gate geometry is shown in Fig.~\ref{f1} a). The current path is
from top to bottom in the figure.
The gates $1$, $2$ and $3$ are
independently controlled. The depression is created by first applying
a negative bias to gates $1$ and $3$ so as to just deplete the
region of 2DEG between the gates. This creates a potential barrier.
A {\it positive} voltage is then applied to gate $2$, thus
forming a depression (or `dimple') in the barrier. 
A schematic cross-section
of the barrier is shown in Fig.~\ref{f1} b). The grid represents
the electrostatic potential of the barrier, including the dimple.
For magnetic fields which are not too small, certain edge states
traverse the barrier, but manifest the shape of the
underlying dimple potential. The heavy lines in Fig.~\ref{f1} b)
represent the edge states at the Fermi energy ($E_F$). We note that
the dimple is capable of supporting localised edge states, as
depicted in the figure.

In Fig.~\ref{f2} we show the experimental results for the conductance of
the dimple structure as a function of applied magnetic field for 
constant side gate voltage and three different dimple gate voltages,
at $50$mK.
The side gate voltage is held constant at $-2.0 V$.
The dimple gate voltages are $+0.025 V$ in $a)$, $+0.15 V$ in
$b)$, and $+0.4 V$ in
$c)$, indicating the evolution of the conductance as the dimple 
becomes wider and deeper. Prominent features are the 
plateau at $2e^2/h$, labeled $P_2$ in the figure, and the development,
with increasing dimple potential, 
of a plateau approaching $4e^2/h$, labeled $P_4$ (particularly for 
{\it b}) and {\it c}).
Notice also the local minima, labeled $M$ and $W$; M
for fields just below the onset of the $P_2$ plateau in
$a)$ (nearly vanishing in $b)$), and W on the low-field side
of the $P_4$ plateau in $b)$ and $c)$. The inset to $b)$ is an 
enlargement of the region around the point $M$. 
In $a)$, the minimum is modulated
by Aharonov-Bohm like oscillations, which persist onto the
plateau $P_2$. As the dimple voltage is increased, the plateau
$P_2$ becomes more pronounced, the wide minimum $M$ shrinks, and
the transition from $P_4$ to $P_2$ becomes sharper (compare
$a)$ and $b)$). The developing wide minimum $W$ for fields
just below the $P_4$ plateau in $b)$ and $c)$ also shows A-B
oscillations, as does the plateau itself. In general, as the dimple evolves,
the conductance develops a progression of spin-unresolved
plateaus ($P_2$ and $P_4$), preceeded by A-B modulated, wide
conductance minima ($M$ and $W$, respectively).
Finally, notice that the high-field
end of the conductance plateau $P_2$ drops off, also showing oscillations,
and that the very low field conductance, while strongly modulated,
increases with dimple voltage.

The nature of the transport problem in the
dimple may be understood
 with the aid of a model based upon the schematic
pictures of Fig.~\ref{f3}. Here the arrowed lines indicate
the edge state configurations analogous to that in 
Fig.~\ref{f1} b). The edge states are assumed to follow the
electrostatic potential, and the 
dimple region supports localised edge states which are allowed due to 
the dimple potential below the center gate itself
(compare Fig.~\ref{f1}). Higher (positive)
dimple voltages widen and deepen the conducting region,
allowing more edge states and/or 
localised states. The edge states and localised states
are coupled together via unitary scattering events, represented 
by the dashed lines in Fig.~\ref{f3}. A feature of the present model 
is that the scattering probabilities are magnetic field-dependent
in a manner that will be discussed below. 

Our theoretical analysis of the above physical picture  is a
generalization of the edge-state scattering theories of 
B\"uttiker,\cite{grge1} Kirczenow and Casta\~no,\cite{grg2}
and calculations in a geometery similar to that considered
here, by Kirczenow;\cite{grg3}
a brief synopsis follows. The calulations are made at $T=0K$.
We assume that the current amplitude
leaving any scattering event is related to the impinging current
amplitude via a unitary scattering matrix. In addition,
the current amplitude acquires a magnetic-field-dependent phase
in transiting the path between scattering events.  
The current amplitude relationships, together
with the unitarity constraint on the scattering matrices,
generate a set of
equations which may be solved for the current amplitudes leaving
the sample in terms of the current amplitudes entering and the
phase accumulated in circulating around 
the localised mode $C$
in Fig.~\ref{f3}. The total transmission of the system (given by 
the square magnitude of the ratio of outgoing current amplitudes
to incoming current amplitudes) may be calculated as a function
of a dimensionless flux, which is the ratio of
the flux threading the closed loop $C$ (or $C'$ as in Fig.~\ref{f3}
c)) to the flux quantum ($h/e$) times $2\pi$, and
is thus a function of both the area of the loop and the applied
field. The conductance thus depends upon the width of the
dimple (and thus the number of modes which traverse), 
as well as the magnetic field. Since exact knowledge of
the potential geometry inside the barrier is impossible, the exact 
diameter of the localised edge state is unknown; however, 
estimates based upon the Aharonov-Bohm periods from Fig.~\ref{f2}
produce reasonable numbers for the sample geometry. For 
instance, an estimated dimple diameter of $500$nm would
give a conversion of roughly $1.3$Telsa for every $10$ in 
units of dimensionless flux. For clarity,
the model calculations are left as a function of dimensionless
flux, with this number as an approximate guide.

Results of the model calculation outlined above are given
in Fig.~\ref{f4}. Here we plot the transmission as a function of
dimensionless flux for three sets of parameters which 
correspond to the different configurations of edge states
shown in Fig.~\ref{f3}. The theory plots in Fig.~\ref{f4}
a), b) and c) should be compared with the experimental
data in Fig~\ref{f2} a), b) and c), respectively, bearing in
mind that the calculations are for $T=0K$, and therefore the
sharp narow features will be smeared out by temperature. We begin with
a qualitative discussion of the physics underlying the plot in
Fig.~\ref{f4} c), 
which will then be extended to the
remaining plots shown in Fig.~\ref{f4}. In general, there are
two main concepts in the model: {\it i}) the number of
edge states decreases with increasing flux, which means that
the higher Landau levels {\it depopulate} as the magnetic field
pushes them through the Fermi energy; {\it ii} at a given value
of flux, the center gate voltage will control the depth and the 
width of the dimple. This will dictate how many localised states
exist in the dimple, as well as how many edge states can traverse
the dimple--generally, the wider (deeper) the dimple is, the more 
edge and localised states it can support. 

In Fig.~\ref{f4} c), we show the total transmission as
a result of 
evolution from the configuration shown in Fig.~\ref{f3} a) to 
Fig.~\ref{f3} b) to
~\ref{f3} c) to ~\ref{f3} d) with increasing magnetic field. A
generic, step-by-step picture of the model is as follows. 
At low fields (see Fig.~\ref{f3} a)), there are three edge states
impinging on the dimple from each direction, and a single 
localised state labeld $C$.
The edge states $B$ and $D$ are coupled to the localised
state via scattering at $1$, $2$, and $4$. As the field is
increased, the state $D$ depopulates, leaving the configuration
shown in Fig.~\ref{f3} b). At higher fields,
the edge state $B$ begins to pinch off and the state $C$
depletes, since
the magnetic field pushes the allowed levels up in energy, eventually
pushing them through the Fermi energy, as mentioned above. This
leads to the situation shown in ~\ref{f3} c), where the edge
state $B$ now couples to the new localised state 
$C'$ (which has developed out of $B$) only through $2$ and $5$, while
edge state $A$ couples to $C'$ at the points of 
maximum curvature, as shown.
From here, the edge state $B$ depopulates, while at the same time
the coupling between $A$ and $C'$ weakens and $C'$ 
depopulates, leaving the situation
depicted in ~\ref{f3} d). Here the edge states $A$ and $G$ are shown 
split into separate spin channels (the subscripts refer to ``up''
and ``down''), since the field splits these in energy. The 
spin channel which is ``against'' the magnetic field will eventually
depopulate. In the model we allow the down edge-states to couple
via cross-channel scattering, which will be stronger for narrow
structures (low dimple voltage). 

In Fig~\ref{f4} c) we begin
at low flux with the configuration of Fig~\ref{f3} a), 
with relatively strong coupling
between states $B,D,E,F$ and $C$, which is 
reasonable for high dimple voltage. The transmission is 
then calculated for
the range of flux from $0$ to $12$ for this configuration and
that in Fig.~\ref{f3} b)--with increasing flux $D$ and $E$
depopulate, and thus coupling
of $D$ and $E$ to $C$ goes smoothly to zero
(which results in the configuration of Fig.~\ref{f3} b)).
The wide conductance minima in the low-flux
regime (an example is labeled $W$ in \ref{f4} c), and in
the experimental trace in Fig.~\ref{f2} c)) are the result of
interference between the closed
loop states composed of state $C$ and the possible loops including
parts of $C$ and parts of $B$ and $F$. For example, in Fig.~\ref{f3}
b) the loop defined by the segments $C1$-$C4$-$F4$-$F2$-$C1$ (here
$C1$ refers to the path $C$ at the site labeled $1$ in Fig.~\ref{f3})
has a different length than the path around loop $C$. The path length
difference, and the corresponding difference in the phase accumulated
in traversing the two paths, can lead to constructive or destructive
interference between the two paths. The minimum $W$ in 
Fig.~\ref{f4} c) is the result of this interference.\cite{ford} 
For the range of
flux $12$ to $30$, we begin with the configuration of Fig~\ref{f3} c),
with strong coupling between all states, and let the scattering
probabilities from  the edge states to the localised state go
smoothly and monotonically to zero as inter-edge-state
scattering is supressed with increasing magnetic field. We
let the scattering from $B$ to $C'$ at $2$ and from $F$ to $C'$ at $5$
vanish more quickly than the others, since the edge state $B$ 
pinches off at lower fields than those which decouple state $A$ from
$C'$. The regular conductance oscillations at $P_4$ 
(compare the inset of Fig.~\ref{f2} c)) are due to
Aharonov-Bohm like interference. 
The drop in conductance from near $4 e^2/h$ at $P_4$ to $2 e^2/h$ 
at $P_2$
is the result of the pinch-off of modes $B,F$, followed by
the decoupling of $A,G$ from mode $C'$. The persistence of
the latter coupling can have interesting effects, as discussed below.
Finally, the oscillatory structure at the high-flux end
(c.f. Fig.~\ref{f2} a)) is due to closed-loop interference
as the spin-down channel pinches off, i.e. the scattering probability
between the spin-down modes across the structure (the dashed
lines in Fig.~\ref{f3} d)) approaches $1$.

The theoretical conductances depicted in Fig~\ref{f4} a) and b) 
are calculated in the same manner as above (ie, the {\it basic
steps are the same}), with some important
differences caused by the lower dimple voltage.
In Fig.~\ref{f4} b), we assume that
the coupling of $D,E$ to $C$ at $2,5$ is weaker than in 
Fig.~\ref{f4} c), consistent with the lower dimple voltage (narrower
structure), while in Fig~\ref{f4} a) we
assume that the initial configuration is given by Fig.~\ref{f3} b);
again the lower dimple voltage allows for fewer edge states initially. 
In Fig.~\ref{f4} a), the lower overall conductance at $P_4$ is
due to weaker coupling between $C'$ and $B,F$ in Fig.~\ref{f3} c), since
here again the structure is narrower and the difference in energy
between levels is greater. 
The minima marked $M$ in Fig.~\ref{f4} a) and b) have the
same origin as the wide minimum marked $W$ in c): after modes 
$B,F$ in Fig.~\ref{f3} c) depopulate, the configuration is
exactly like that of Fig.~\ref{f3} b), with one less edge state.
The resulting oscillatory structure at
$M$ is due to the persistent coupling of
$A,G$ to $C'$ as mentioned above. This is especially apparent in
Figs.~\ref{f4} a) and \ref{f2} a), but is also visible in 
\ref{f4} b) and \ref{f2} b). 

In general, comparing the experimental and theoretical pictures
reveals the following. A developing conductance plateau $P$ is 
preceded on the low-field side by a local minimum $M$
or $W$, which
is modulated by Aharonov-Bohm like oscillations. These minima
are the result of interference between edge states and
localized states in the dimple--compare the features marked $M$
and $W$
in Fig.~\ref{f2} and Fig.~\ref{f4}. Also, more plateaus
develop with higher dimple voltage as more edge states are 
allowed. 
The evolution of the small plateau near
$4 e^2/h$ (labeled $P_4$) in the experimental data is a result of 
the wider dimple
created by higher gate voltage. In the model, this widening has two
consequences: the  inital coupling of $B$ and $F$ 
to $C$ in the configuration of Fig~\ref{f3} c) is stronger, and
the couplings between modes decay more rapidly with increasing flux
(a consequence of the dimensionless flux having the area of
the dimple biult-in. The field required to de-populate levels 
goes down as area increases).
The overall effect is the
appearance of a hump ($P_4$), and then a short plateau--compare Fig~\ref{f2}
b) and c), comparing the labeled features. The short plateau 
$P_4$ is still modulated by A-B oscillations,
which are the result of the remaining coupling and interference. Note
that $P_4$ narrows with increasing dimple voltage in both 
experiment and theory.

In summary, we have presented an experimental study of the
magneto-conductance of a dimple, a structure formed by 
gradually imposing a sub-micron scale depression in a potential barrier
across a 2DEG. The role of the dimpled potential barrier
in the backscattering of magnetic edge states
and the quantum Hall effect is studied in
a controlled fashion. We find that scattering and interference
between edge states and
localised states within the dimple produces a set of broad
minima and oscillations in the conductance as a function of
magnetic field.

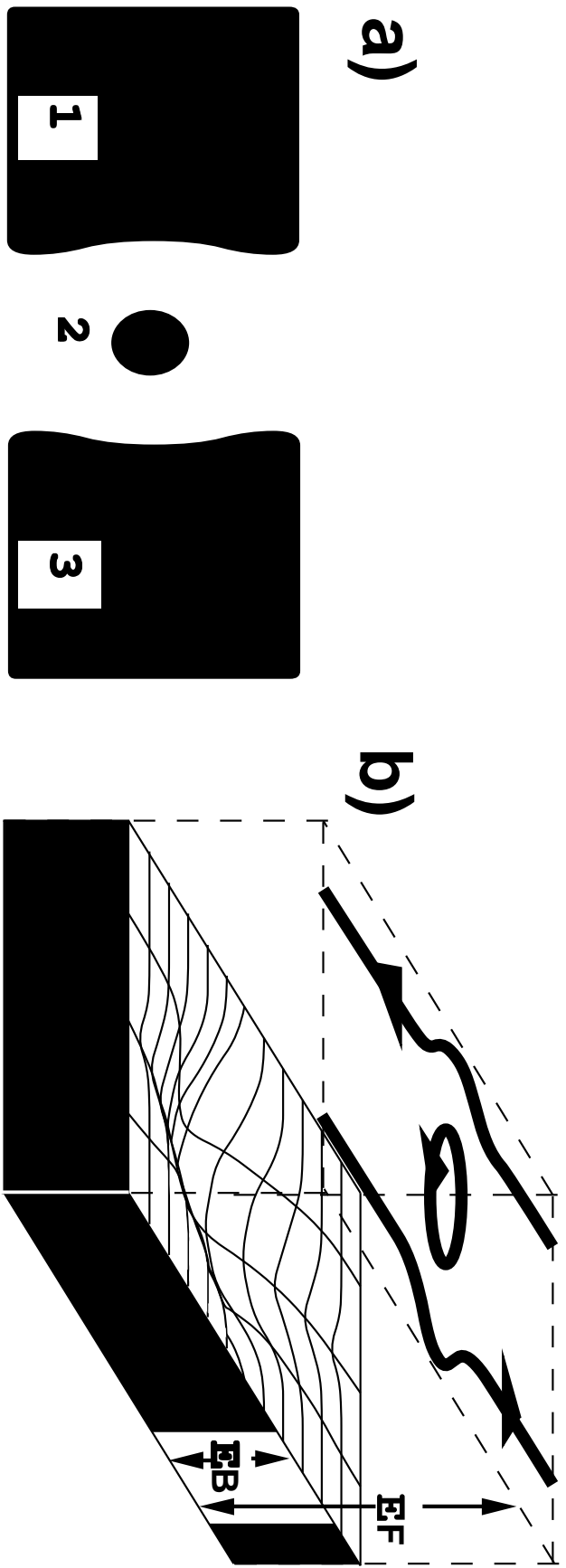
\begin{figure}
\caption{a) Schematic of the experimental geometry. The dark
areas represent the gates. All gates are individually 
controlled. For the experiments reported here, gates
$1$ and $3$ are biased with negative voltage
creating a potential barrier, then a positive bias is
applied to the center gate $2$, creating the dimple
in the potential barrier. The gate $2$ has a diameter of 
$300$nm, and gates $1$ and $3$ are separated by $850$ nm.
The GaAs/AlGaAs wafer was delta doped with a Si
donor density of $1.5X10^{12} cm^{-2}$ separated from the
2DEG by a $400\AA$ spacer layer. The density and mobility
after illumination were $3.3X10^{11} cm^{-2}$ and
$1.5X10^6 cm/Vs$, respectively. The gates are made
from $Ti/Pt/Au$ alloy. b)Schematic of the 
profile (grid) and the resulting edge states. 
For reference, the barrier
height ($E_B$) and the Fermi level ($E_F$) 
are shown.}\label{f1}
\end{figure}
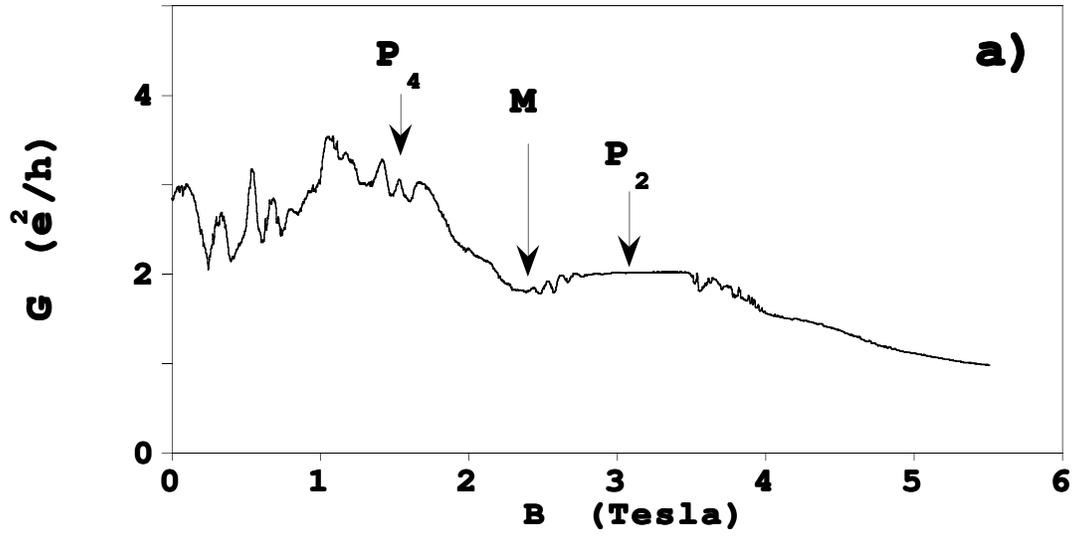
\begin{figure}
\caption{Experimental results for the conductance as
a function of magnetic field for three different center
gate voltages. Here the side gates are maintained at 
$-2 V$, while the center gate is varied: a)$0.025 V$,
b)$0.15 V$ and c)$0.4 V$.}\label{f2}
\end{figure}
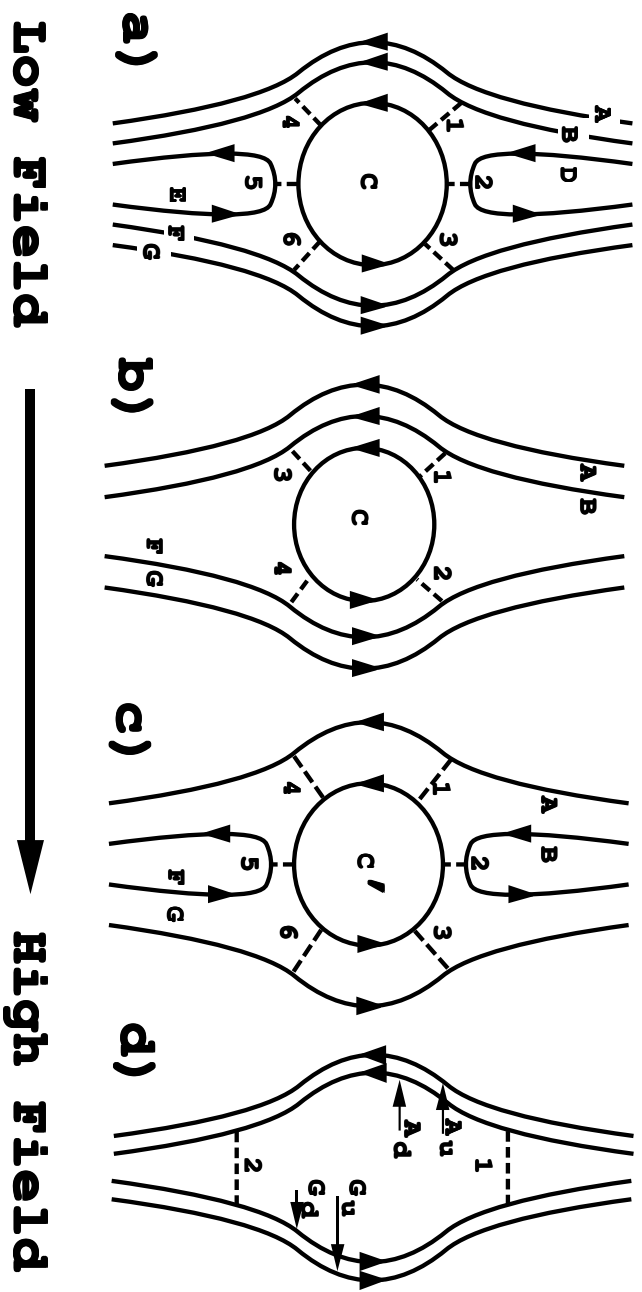
\begin{figure}
\caption{Model configurations considered in this paper. Here
the arrowed lines are edge states (compare Fig.~\protect\ref{f1} b)),
and the dashed lines
represent inter-state scattering. The labels are defined in the text.
In c), the subscripts refer to spin.}\label{f3}
\end{figure}
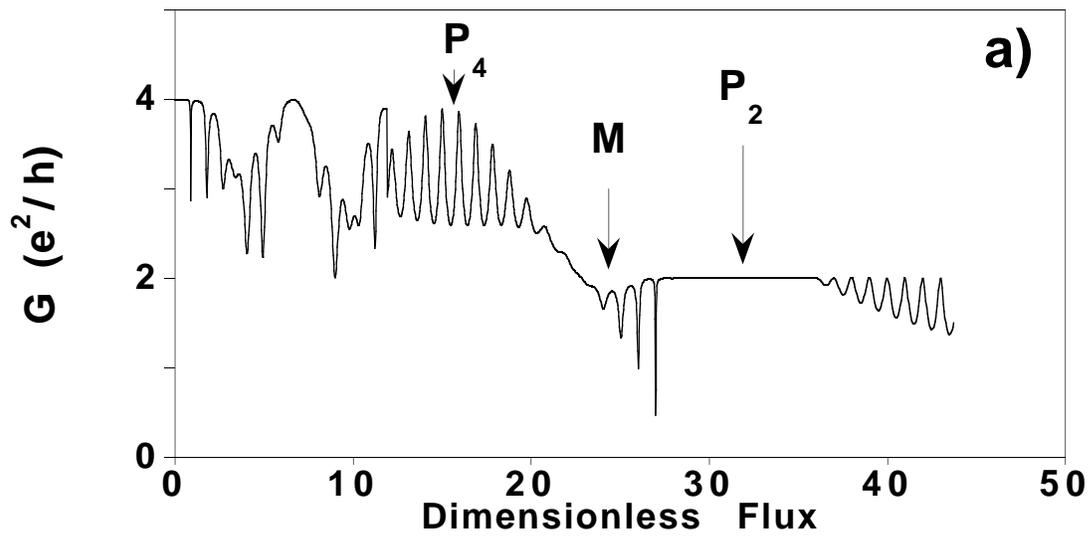
\begin{figure}
\caption{Results of the model calculations described in the text.
Note the labeled features for comparison to the experimental
data of Fig.~\protect\ref{f2}. The dimensionless flux is 
given by the ratio of the magnetix flux through the closed loop
paths of Fig.~\protect\ref{f3} to the flux quantum times $2\pi$,
as discussed in the text.}\label{f4}
\end{figure}
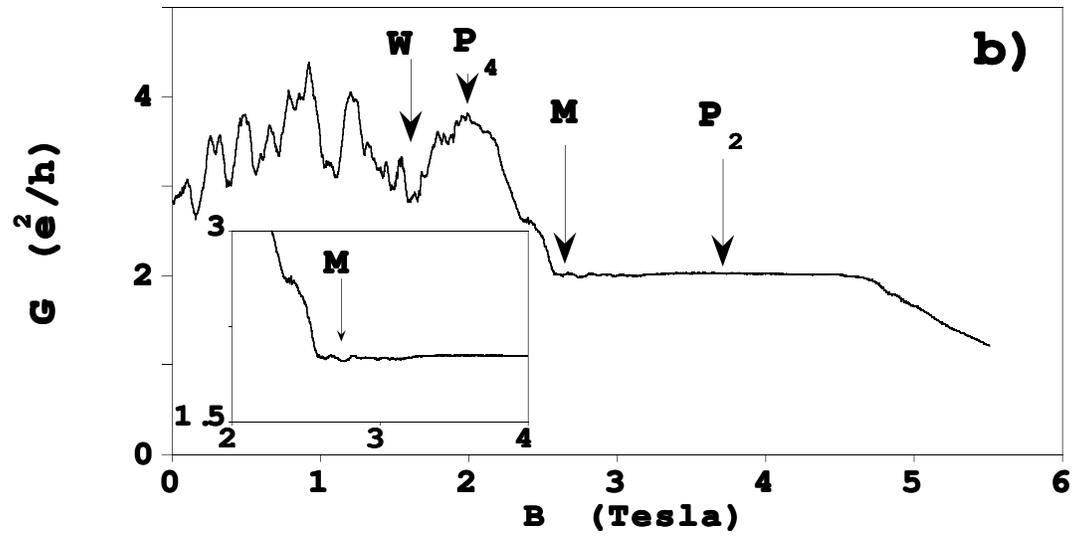
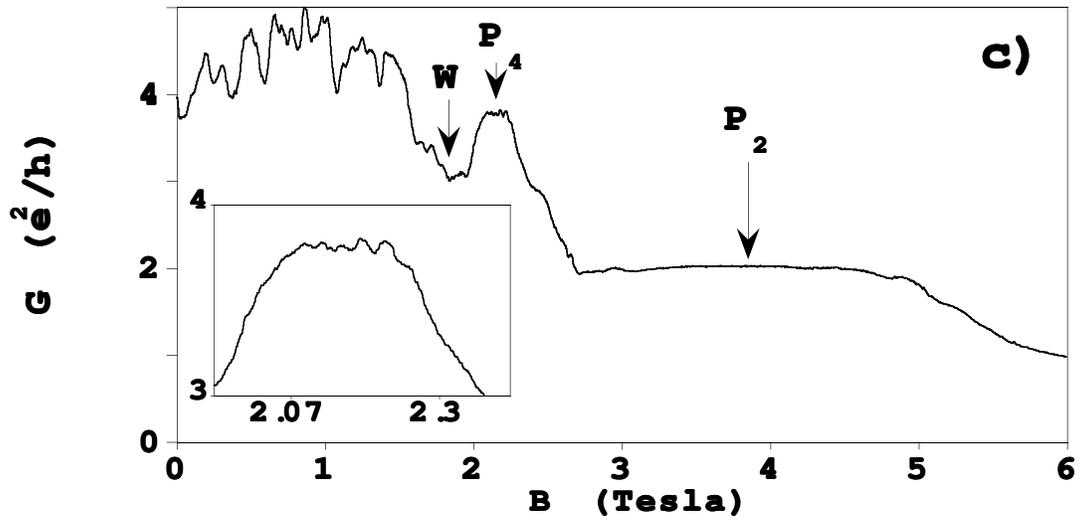
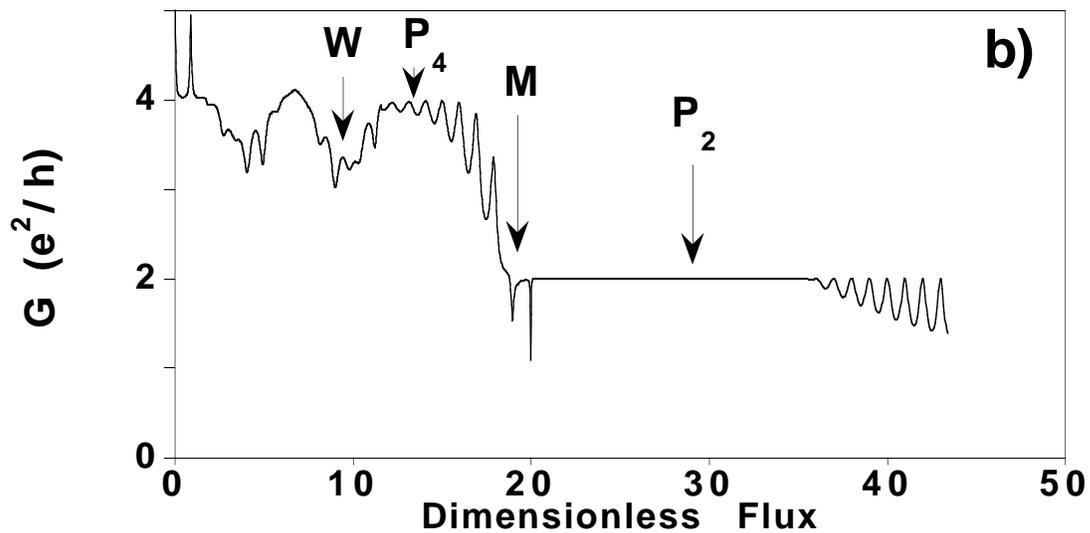
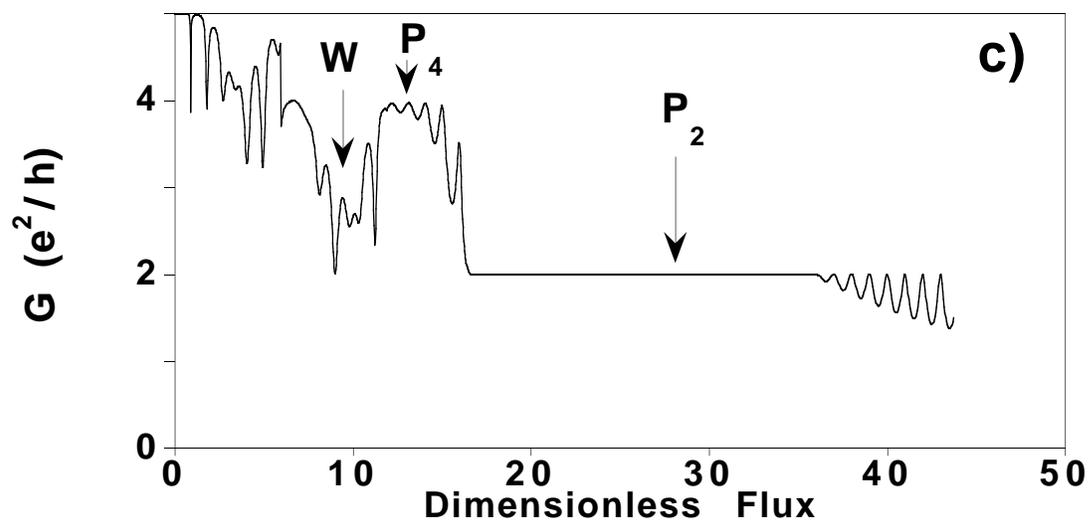
\end{document}